

\message
{JNL.TEX version 0.92 as of 6/9/87.  Report bugs and problems to Doug Eardley.}

\catcode`@=11
\expandafter\ifx\csname inp@t\endcsname\relax\let\inp@t=\input
\def\input#1 {\expandafter\ifx\csname #1IsLoaded\endcsname\relax
\inp@t#1%
\expandafter\def\csname #1IsLoaded\endcsname{(#1 was previously loaded)}
\else\message{\csname #1IsLoaded\endcsname}\fi}\fi
\catcode`@=12







\def\beginlinemode{\endmode
  \begingroup\parskip=0pt \obeylines\def\\{\par}\def\endmode{\par\endgroup}}
\def\beginparmode{\endmode
  \begingroup \def\endmode{\par\endgroup}}
\let\endmode=\par
{\obeylines\gdef\
{}}
\def\singlespace{\baselineskip=\normalbaselineskip}

\def\oneandahalfspace{\baselineskip=\normalbaselineskip
  \multiply\baselineskip by 3 \divide\baselineskip by 2}
\def\doublespace{\baselineskip=\normalbaselineskip \multiply\baselineskip by 2}






\def\title			
  {\null\vskip 3pt plus 0.2fill
   \beginlinemode \doublespace \raggedcenter \bf}

\def\author			
  {\vskip 3pt plus 0.2fill \beginlinemode
   \singlespace \raggedcenter\sc}

\def\affil			
  {\vskip 3pt plus 0.1fill \beginlinemode
   \oneandahalfspace \raggedcenter \sl}

\def\abstract			
  {\vskip 3pt plus 0.3fill \beginparmode
   \oneandahalfspace ABSTRACT: }

\def\endtitlepage		
  {\endpage			
   \body}

\def\body			
  {\beginparmode}		

\def\beginitems{
\par\medskip\bgroup\def\i##1 {\item{##1}}\def\ii##1 {\itemitem{##1}}
\leftskip=36pt\parskip=0pt}
\def\enditems{\par\egroup}

\def\beneathrel#1\under#2{\mathrel{\mathop{#2}\limits_{#1}}}

\def\refto#1{~[{#1}]}

\def\references			
  {
   \beginparmode
   \frenchspacing \parindent=0pt \leftskip=1truecm
   \parskip=8pt plus 3pt \everypar{\hangindent=\parindent}}

\gdef\refis#1{\item{#1.\ }}			

\gdef\journal#1, #2, #3, 1#4#5#6{		
    {\sl #1~}{\bf #2}, #3 (1#4#5#6)}		

\def\endreferences{\body}

\catcode`@=11
\newcount\r@fcount \r@fcount=0
\newcount\r@fcurr
\immediate\newwrite\reffile
\newif\ifr@ffile\r@ffilefalse
\def\w@rnwrite#1{\ifr@ffile\immediate\write\reffile{#1}\fi\message{#1}}

\def\writer@f#1>>{}
\def\referencefile{
  \r@ffiletrue\immediate\openout\reffile=\jobname.ref%
  \def\writer@f##1>>{\ifr@ffile\immediate\write\reffile%
    {\noexpand\refis{##1} = \csname r@fnum##1\endcsname = %
     \expandafter\expandafter\expandafter\strip@t\expandafter%
     \meaning\csname r@ftext\csname r@fnum##1\endcsname\endcsname}\fi}%
  \def\strip@t##1>>{}}

\def\citeall#1{\xdef#1##1{#1{\noexpand\cite{##1}}}}
\def\cite#1{\each@rg\citer@nge{#1}}	

\def\each@rg#1#2{{\let\thecsname=#1\expandafter\first@rg#2,\end,}}
\def\first@rg#1,{\thecsname{#1}\apply@rg}	
\def\apply@rg#1,{\ifx\end#1\let\next=\relax
\else,\thecsname{#1}\let\next=\apply@rg\fi\next}

\def\citer@nge#1{\citedor@nge#1-\end-}	
\def\citer@ngeat#1\end-{#1}
\def\citedor@nge#1-#2-{\ifx\end#2\r@featspace#1 
  \else\citel@@p{#1}{#2}\citer@ngeat\fi}	
\def\citel@@p#1#2{\ifnum#1>#2{\errmessage{Reference range #1-#2\space is bad.}%
    \errhelp{If you cite a series of references by the notation M-N, then M and
    N must be integers, and N must be greater than or equal to M.}}\else%
 {\count0=#1\count1=#2\advance\count1 by1\relax\expandafter\r@fcite\the\count0,%
  \loop\advance\count0 by1\relax
    \ifnum\count0<\count1,\expandafter\r@fcite\the\count0,%
  \repeat}\fi}

\def\r@featspace#1#2 {\r@fcite#1#2,}	
\def\r@fcite#1,{\ifuncit@d{#1}
    \newr@f{#1}%
    \expandafter\gdef\csname r@ftext\number\r@fcount\endcsname%
                     {\message{Reference #1 to be supplied.}%
                      \writer@f#1>>#1 to be supplied.\par}%
 \fi%
 \csname r@fnum#1\endcsname}
\def\ifuncit@d#1{\expandafter\ifx\csname r@fnum#1\endcsname\relax}%
\def\newr@f#1{\global\advance\r@fcount by1%
    \expandafter\xdef\csname r@fnum#1\endcsname{\number\r@fcount}}

\let\r@fis=\refis			
\def\refis#1#2#3\par{\ifuncit@d{#1}
   \newr@f{#1}%
   \w@rnwrite{Reference #1=\number\r@fcount\space is not cited up to now.}\fi%
  \expandafter\gdef\csname r@ftext\csname r@fnum#1\endcsname\endcsname%
  {\writer@f#1>>#2#3\par}}

\def\ignoreuncited{
   \def\refis##1##2##3\par{\ifuncit@d{##1}%
     \else\expandafter\gdef\csname r@ftext\csname r@fnum##1\endcsname\endcsname%
     {\writer@f##1>>##2##3\par}\fi}}

\def\r@ferr{\endreferences\errmessage{I was expecting to see
\noexpand\endreferences before now;  I have inserted it here.}}
\let\r@ferences=\references
\def\references{\r@ferences\def\endmode{\r@ferr\par\endgroup}}

\let\endr@ferences=\endreferences
\def\endreferences{\r@fcurr=0
  {\loop\ifnum\r@fcurr<\r@fcount
    \advance\r@fcurr by 1\relax\expandafter\r@fis\expandafter{\number\r@fcurr}%
    \csname r@ftext\number\r@fcurr\endcsname%
  \repeat}\gdef\r@ferr{}\endr@ferences}


\let\r@fend=\endpaper\gdef\endpaper{\ifr@ffile
\immediate\write16{Cross References written on []\jobname.REF.}\fi\r@fend}

\catcode`@=12

\citeall\refto		

\magnification 1200
\vsize=7.8in
\hsize=5.3in
\voffset=0.1in
\hoffset=-0.05in
\newcount\eqnumber
\baselineskip 18pt plus 0pt minus 0pt


\font\rmmthree=cmbx10 scaled 1500
\font\rmmtwo=cmbx10 scaled 1200
\font\rmmoneB=cmbx10 scaled 1100

\font\rmmoneI=cmti10 scaled 1000

\font\ninerm=cmr10 scaled 900

\font\eightit=cmti10 scaled 800
\font\eightrm=cmr10 scaled 800
\font\eightbf=cmbx10 scaled 800


\def\title#1{\centerline{\noindent{\rmmthree #1}}\nobreak\medskip\eqnumber=1}

\def\sectbegin#1#2{\bigskip\bigbreak\leftline{\rmmtwo
#1~~#2}\nobreak\medskip\nobreak}

\def\nosectbegin#1{\bigskip\bigbreak\leftline{\rmmtwo #1}\nobreak\medskip}


\def\lapp{\hbox{$ {     \lower.40ex\hbox{$<$}
                   \atop \raise.20ex\hbox{$\sim$}
                   }     $}  }
\def\gapp{\hbox{$ {     \lower.40ex\hbox{$>$}
                   \atop \raise.20ex\hbox{$\sim$}
                   }     $}  }

\def\marbul{\strut\vadjust{\kern-2pt$\bullet$}}

\def\specialwarn{\vtop to
\strutdepth{\baselineskip\strutdepth\vss\llap{
\lower.1ex\hbox{$\bigtriangleup$}\kern-0.884em$\triangle$\kern-0.5667em{\eightrm
!}\hskip 13.5pt}\null}}
\def\strutdepth{\dp\strutbox}


\def\new{{\the\eqnumber}\global\advance\eqnumber by 1}
\def\delaynew{{\the\eqnumber}}
\def\nownew{\global\advance\eqnumber by 1}
\def\last{\advance\eqnumber by -1 {\the\eqnumber}
    \global\advance\eqnumber by 1}
\def\eqnam#1{
\xdef#1{\the\eqnumber}}


\def\figure#1#2#3#4#5{
\topinsert
\null
\medskip
\vskip #2\relax
\null\hskip #3\relax
\special{illustration #1}
\medskip
{\baselineskip 10pt\noindent\narrower\rm\hbox{\eightbf
Figure #4}:\quad\eightrm
#5 \smallskip}
\endinsert}

\def\vdoublefigure#1#2#3#4#5#6#7#8{
\topinsert
\medskip
\null
\vskip #3\relax
\null\hskip #4\relax
\special{illustration #1}
\smallskip
\vskip #6\relax
\line{\null\hskip #5\relax
\special{illustration #2}\hfill}
\medskip
{\baselineskip 10pt\noindent\narrower\rm\hbox{\eightbf
Figure #7}:\quad\eightrm
#8 \smallskip}
\endinsert}

\def\caption#1#2{
\baselineskip 10pt\noindent\narrower\rm\hbox{\eightbf
#1}:\quad\eightrm
#2 \smallskip}

\def\picture #1 by #2 (#3){
  \vbox to #2{
    \hrule width #1 height 0pt depth 0pt
    \vfill
    \special{picture #3} 
    }
  }

\def\scaledpicture #1 by #2 (#3 scaled #4){{
  \dimen0=#1 \dimen1=#2
  \divide\dimen0 by 1000 \multiply\dimen0 by #4
  \divide\dimen1 by 1000 \multiply\dimen1 by #4
  \picture \dimen0 by \dimen1 (#3 scaled #4)}
  }


\def\fa{f_{\rm a}}

\def\dalemb#1#2{{\vbox{\hrule height .#2pt
\hbox{\vrule width.#2pt height#1pt \kern#1pt\vrule width.#2pt}
\hrule height.#2pt}}}

\def\tdot{\kern -8.5pt {}^{{}^{\hbox{...}}}}
\def\dotprime{\kern -8.0pt{}^{{}^{\hbox{.}~\prime}}}


\title{STRING RADIATIVE BACKREACTION}

\medskip
\centerline{\rmmoneB R.$\,$A. Battye {\rmmoneI~and~} E.$\,$P.$\,$S. Shellard}
\vskip 6pt
\baselineskip 12pt
\centerline{\rmmoneI Department of Applied Mathematics and Theoretical Physics}

\centerline{\rmmoneI University of Cambridge}

\centerline{{\rmmoneI Silver Street, Cambridge~CB3 9EW, U.K.}
\footnote{*}{{\noindent Email: rab17$\,$@$\,$amtp.cam.ac.uk and
epss$\,$@$\,$amtp.cam.ac.uk}
\smallskip\indent Paper
submitted to {\eightit Phys. Rev. Letters}.\smallskip}}

\smallskip
\centerline{\rmmoneI and}
\smallskip
\centerline{\rmmoneI Isaac Newton Institute for Mathematical Sciences}
\centerline{\rmmoneI Clarkson Road, Cambridge~CB3 0EH, U.K.}

\bigskip \centerline{\rmmoneB Abstract} \medskip {\narrower{\baselineskip 9pt
\ninerm \noindent We discuss radiative backreaction for global strings
described by the Kalb-Ramond action with an analogous derivation to that for 
the point electron in classical electrodynamics. We show how local corrections
to the equations of motion allow one to separate the self-field of the string
from that of the radiation field.  Modifications to this `local backreaction
approximation' circumvent the runaway solutions familiar from electrodynamics,
allowing these corrections to be used to evolve string trajectories
numerically. Comparisons are made with analytic and numerical radiation
calculations from previous work and the merits and limitations of this
approach are discussed. Finally, we comment on the relevance of this work to
simulations of a cosmological network of global or axion strings. These
methods can also be applied to describe gravitational, electromagnetic and
other forms of string radiative backreaction. \smallskip}}  

\smallskip \baselineskip 12pt\tenrm

\sectbegin{1~}{Introduction}

\noindent Global strings arise naturally in theories with a spontaneously
broken global $U(1)$ symmetries, such as axion models, superstrings and some
GUT models.  The Goldstone bosons or axions, associated with the broken
symmetry, can be radiated by an oscillating global string.   This radiation
power can be calculated using perturbed trajectories, but the effect of
radiative damping on the actual string motion remains to be adequately
understood.  This deficiency is a major obstacle impeding the quantitative
description of string network evolution.  Radiative backreaction is believed to
determine the nature of long string small-scale structure and small loop
creation sizes and, hence, the amplitude and spectrum of the resulting
radiation background.

It is appropriate to begin discussion of the problem of radiative backreaction
by considering the point electron in classical relativistic
electrodynamics\refto{dirac,barut,jackson}. In this case, the self and
radiation fields of the electron can be distinguished easily since the
self-field is of order $1/R^2$, whereas the radiation field fall-off is of
order $1/R$. Careful analysis of the equations of motion leads to the
renormalisation of the electron mass $M_{\rm ren} = \textstyle{4\over 3}
M_{\rm e}$ by the Coulomb self-field and the first-order approximation to the
radiation force---the Abraham--Lorentz force---is given by
\eqnam{\abrahamlorentz}  $$ F_{\mu}^{\rm rad}= -{2\over 3}~{e^2\over
4\pi}\bigg{(}X^{\tdot}_{\mu}+\ddot X^2 \dot X_{\mu}\bigg{)}\,,\eqno(\new) $$ 
where $X_{\mu}(\tau)$ is the position on the electron's worldline at time
$\tau$. However, the dependence of this force on $X^{\tdot}_{\mu}$ leads to
problems in numerical applications because (\abrahamlorentz) has exponentially
increasing solutions. These unphysical `runaway' solutions can only be
suppressed by rewriting the equations of motion as an integro-differential
equation.   In the following we shall derive the analogue of (\abrahamlorentz)
for a global string using an antisymmetric tensor formalism.

\sectbegin{2~}{Self-field renormalisation}

\noindent The essential features of global strings in flat space are exhibited
in the simple U(1) Goldstone model, with action given by \eqnam{\goldaction} 
$$ S=\int d^4x\bigg{\{}\partial_{\mu}\bar\Phi\partial^{\mu}\Phi-{1\over
4}\lambda(\bar\Phi\Phi-\fa^2)^2\bigg{\}}\,,\eqno(\new) $$ where $\Phi$ is a
complex scalar field which can be split into a massive (real) component $\phi$
and a massless (real periodic) Goldstone boson $\vartheta$. The analytic
treatment of global string dynamics is hampered by the topological coupling of
the self field of the string to the Goldstone boson radiation field. However,
we can exploit the well-known duality between a massless scalar field and a
two-index antisymmetric tensor $B_{\alpha\beta}$ to replace the Goldstone
boson $\vartheta$ in (\goldaction) via the relation
$\phi^2\partial_{\mu}\vartheta = {1\over
2}\fa\epsilon_{\mu\nu\lambda\rho}\partial^{\mu}B^{\lambda\rho}$. Performing
this transformation carefully and integrating over the massive degrees of
freedom about the two-dimensional string worldsheet $X^\mu(\sigma,
\tau)$\refto{DSb},  yields the flat-space Kalb--Ramond
action\refto{KR,Wit,VV}, \eqnam{\kraction}   
$$  
S  = -\mu_0\int\nolimits
\sqrt{-\gamma} \, d\sigma d\tau - {1 \over 6}\int\nolimits\sqrt{-g}\, d^4x\, H^2
- 2\pi f_a\int\nolimits B_{\alpha\beta}V^{\alpha\beta}d\sigma
d\tau\,,\eqno(\new) 
$$  
where $H_{\mu\alpha\beta}
=\partial_{\mu}B_{\alpha\beta} +\partial_{\beta}B_{\mu\alpha}+
\partial_{\alpha}B_{\beta\mu}$ is the field strength of $B_{\alpha\beta}$, the
worldsheet metric is $\gamma_{ab}= g_{\mu\nu} \partial_a X^\mu \partial_b
X^\nu$ with $\gamma=\det(\gamma_{ab})$, and $V_{\alpha\beta}=
\epsilon^{ab}\partial_a X_\alpha \partial_b X_\beta$ is the antisymmetric
vertex operator.

Varying the action (\kraction) with respect to the worldsheet coordinates and
the antisymmetric tensor  yields the string equations of motion and 
the tensor field equations,   \eqnam{\fulleom} $$
\eqalign{&\mu_0\partial_a\big{(}\sqrt{-\gamma}\gamma^{ab}\partial_bX^{\mu}\big{)}
 = F^{\mu} =  2\pi\fa H^{\mu\alpha\beta}V_{\alpha\beta}\,,\cr
&\partial_{\mu}H^{\mu\alpha\beta}=4\pi J^{\alpha\beta}=2\pi\fa\int d\sigma
d\tau\,\delta^{4}\big{(}
x-X(\sigma,\tau)\big{)}\,V^{\alpha\beta}\,.}\eqno(\new)$$ In the Lorentz gauge,
$\partial_{\mu}B^{\mu\nu}=0$, the field equations can be solved using standard
Green's function methods in retarded time, \eqnam{\bsoln} $$B_{\alpha\beta}(x)=
2\pi\fa\int d\bar\sigma d\bar\tau \, D_{\rm
ret}(x-X(\bar\sigma,\bar\tau))V_{\alpha\beta}(\bar\sigma,\bar\tau)\,,\eqno(\new)$$
where $D_{\rm ret}(x)= (1/2\pi)\theta(x_0)\delta(x^2)$ is the retarded Green's
function. Defining $\Delta_{\mu}=x_{\mu}-X_{\mu}(\bar\sigma,\bar\tau)$ and
treating $\bar\sigma=\bar\sigma(\bar\tau)$ one can perform various standard
manipulations\refto{barut,DQ,HCH} to obtain the Lienard-Wiechart potential and
its derivative \eqnam{\lwpot} $$ B_{\alpha\beta}(x)= {\fa\over 2}\int
d\bar\sigma  \bigg{(}{V_{\alpha\beta}\over |\Delta.\dot
X|}\bigg{)}\bigg{|}_{\bar\tau=\tau_R}\,,\quad\partial_{\mu}B_{\alpha\beta}(x)=
{\fa\over 2}\int d\bar\sigma  {1\over \Delta.\dot X}{\partial\over
\partial\bar\tau}\bigg{(}{\Delta_{\mu}V_{\alpha\beta}\over |\Delta.\dot
X|}\bigg{)}\bigg{|}_{\bar\tau=\tau_R}\,,\eqno(\new) $$ where
$\Delta^2|_{\bar\tau=\tau_R}=0$ and $\tau_R<t$.  One can separate the
radiation field from the self field by noting that the radiation field is a
solution of the homogeneous wave equation. If one defines the  advanced Green's
function, $D_{\rm adv}(x)=(1/2\pi)\theta(-x_0)\delta(x^2)$, then by analogy
with electrodynamics\refto{barut,DQ} one can define Green's functions for the
self and radiation fields $2D_{\rm self}(x)=D_{\rm ret}(x)+D_{\rm adv}(x)$ and
$2D_{\rm rad}(x)=D_{\rm ret}(x)-D_{\rm adv}(x)$. Thus we can calculate the
derivatives of the Lienard-Wiechart potentials for both components
\eqnam{\lwpotselfrad}  $$ \eqalign{\partial_{\mu}B_{\alpha\beta}^{\rm
self}(x)&={\fa\over 4}\int d\bar\sigma  \bigg{[}{1\over \Delta.\dot X}
{\partial\over \partial\bar\tau}\bigg{(}{\Delta_{\mu}V_{\alpha\beta}\over
|\Delta.\dot X|}\bigg{)}\bigg{|}_{\bar\tau=\tau_R}+{1\over \Delta.\dot X}
{\partial\over \partial\bar\tau}\bigg{(}{\Delta_{\mu}V_{\alpha\beta}\over
|\Delta.\dot X|}\bigg{)}\bigg{|}_{\bar\tau=\tau^{\prime}_R}\bigg{]}\,,\cr
\partial_{\mu}B_{\alpha\beta}^{\rm rad}(x)&={\fa\over 4}\int d\bar\sigma 
\bigg{[}{1\over \Delta.\dot X} {\partial\over
\partial\bar\tau}\bigg{(}{\Delta_{\mu}V_{\alpha\beta}\over |\Delta.\dot
X|}\bigg{)}\bigg{|}_{\bar\tau=\tau_R}-{1\over \Delta.\dot X} {\partial\over
\partial\bar\tau}\bigg{(}{\Delta_{\mu}V_{\alpha\beta}\over |\Delta.\dot
X|}\bigg{)}\bigg{|}_{\bar\tau=\tau^{\prime}_R}\bigg{]}\,. }\eqno(\new) $$ where
$\Delta^2|_{\bar\tau=\tau_R,\tau^{\prime}_R}=0$ and
$\tau_R<t,\tau^{\prime}_R>t$. 

The equations of motion (\fulleom) are problematic because the field diverges
as any point on the string is approached $x_{\mu}\rightarrow
X_{\mu}(\sigma,\tau)$, as for the point electron in classical electrodynamics.
The electron has a natural scale, the classical electron radius, which allows
the renormalisation of the electron mass and the calculation of the
first-order approximation to the radiation force (\abrahamlorentz). In a
similar manner for strings, the short distance divergences are cut off by the
natural scale of the string width $\delta$.  However, the string self-field 
also diverges at large distances and so we must introduce another arbitrary
cut-off scale $\Delta$.  Formally, this scale corresponds to the integrals in
(\bsoln), (\lwpot) and (\lwpotselfrad) running over a region of size $\Delta$
around the point in question, that is, we only consider the effects from
neighbouring string points within a distance of $\Delta/2$. Note that we
define $\Delta$ in terms of an invariant length of string.
Physically, it will correspond to a distance beyond which the long-range
string fields become uncorrelated and begin to cancel, so we are
essentially approximating the effect of the nearby string with a square or `top hat'
window function.   Our expectation, then, is that an appropriate choice for $\Delta$
would be near the average curvature radius of the string.

For flat-space string dynamics, the conformal string gauge is usually
employed. However, when considering problems in which the string energy decays
it is best to use the temporal transverse gauge in which $X^0=t=\tau$ and 
$\dot {\bf X}.{\bf X}^{\prime}=0$ with $X^\mu=(t,{\bf X})$.  The equations of
motion for the string (\fulleom) and the energy of the string become
\eqnam{\tempeom}  $$ \mu_0\bigg{(}\ddot{\bf X}-{1\over \epsilon}\bigg{(}{{\bf
X}^{\prime}\over\epsilon}\bigg{)}^{\prime}\bigg{)} = {\bf f}\,,\quad
\mu_0\dot\epsilon = f^0\,,\quad E=\mu_0\int d\sigma\epsilon\,,\eqno(\new) $$ 
where $\epsilon^2 = \dot {\bf X}^2/(1-{\bf X}^{\prime 2})$ and
$F^{\mu}=(f^0,{\bf f})$. 

We will now renormalize the equations of motion (\tempeom), though the details
of the procedure will be described elsewhere\refto{BSa} (see also
refs.\refto{DQ,HCH,BS}). For the purposes of this paper, it suffices to note that the
expression for the derivatives of the Lienard-Wiechart potentials of the self and
radiation fields (\lwpotselfrad) can be expanded locally about some point on the
string. As discussed above, this procedure strongly suggests a natural scale for the
otherwise arbitrary renormalisation cut-off $\Delta$ near the average curvature
radius of the string. After a
detailed set of manipulations which involve use of the gauge conditions, one can
deduce that \eqnam{\renfields} $$ 
\eqalign{ H_{\mu\alpha\beta}^{\rm self}=&{\fa\over 2\dot
X^4}\bigg{[}{1\over \epsilon}\ddot X_{[\rho}V_{\alpha\beta]} -
{1\over\epsilon^3}X^{\prime\prime}_{[\rho}V_{\alpha\beta]}\bigg{]}\log(\Delta/\delta)+{\cal
O} (\Delta^2)\,,\cr H_{\mu\alpha\beta}^{\rm rad}=&{\fa\over 2\dot
X^4}\bigg{[}-{4\over 3}X^{\tdot}_{[\rho}V_{\alpha\beta]}-{1\over 2}\dot
X_{[\rho}\dot V_{\alpha\beta]}+{3\dot\epsilon\over
2\epsilon}X^{\prime\prime}_{[\rho}V_{\alpha\beta]}\cr&+\bigg{(} {2\dot X.\ddot
X\over \dot X^2} - {\dot\epsilon\over 2\epsilon}\bigg{)}\ddot
X_{[\rho}V_{\alpha\beta]}\bigg{]}\Delta+{\cal O} (\Delta^2)\,,\cr}\eqno(\new)
$$  
where
$A_{[\mu\alpha\beta]}=A_{\mu\alpha\beta}+A_{\beta\mu\alpha}+A_{\alpha\beta\mu}$.
Note that the self-field has no order $\Delta$ term. Ignoring terms of order
$\Delta^2$, we can then deduce expressions for the self-force and the radiation
backreaction force,   \eqnam{\tempforce}   
$$
\eqalign{{\bf f}^{\rm self} =& -2\pi\fa^2\log(\Delta/\delta)\bigg{(}\ddot{\bf
X}-{1\over\epsilon}\bigg{(}{{\bf
X}^{\prime}\over\epsilon}\bigg{)}^{\prime}\bigg{)}\,,\cr 
{\bf f}^{\rm rad} =
& \pi\fa^2\Delta\bigg{\{}{4\over 3}\epsilon{\bf X}^{\tdot}
+\bigg{[}2\epsilon\bigg{(}{\dot{\bf X}\cdot\ddot{\bf X}\over 1-\dot{\bf X}^2}
\bigg{)} + 3\dot\epsilon\bigg{]}\ddot{\bf X} - {2\over\epsilon}\bigg{(} {{\bf
X}^{\prime}\cdot\ddot{\bf X}\over 1-\dot{\bf X}^2}\bigg{)}{\dot{\bf X}}^{\prime}
-{3\dot\epsilon\over\epsilon^2}{\bf X}^{\prime\prime}\cr 
+& \bigg{[}-{4\over
3\epsilon}\bigg{(}{{\bf X}^{\prime}\cdot {\bf X}^{\tdot}\over 1-\dot{\bf
X}^2}\bigg{)} - {4\over\epsilon}{(\dot{\bf X}\cdot\ddot{\bf X} )({\bf
X}^{\prime}\cdot\ddot{\bf X})\over (1-\dot{\bf X}^2)^2} - {\dot\epsilon
\over\epsilon^2}\bigg{(}{{\bf X}^{\prime}\cdot\ddot{\bf X}\over 1 -\dot{\bf X}^2}
\bigg{)}+{3\dot\epsilon\over\epsilon^4}\bigg{(}{\dot{\bf X}\cdot{\bf X}^{\prime
\prime}\over 1 -\dot{\bf X}^2}\bigg{)}\bigg{]}{\bf X}^{\prime}\bigg{\}} \,,\cr
f^{0,\rm self}=& -2\pi\fa^2\log(\Delta/\delta)\dot\epsilon\,,\cr
 f^{0,\rm
rad}=& \pi\fa^2\Delta\bigg{\{}{4\over 3}\epsilon^2\bigg{(}{\dot{\bf X}\cdot
{\bf X}^{\tdot}\over 1-\dot {\bf X}^2}\bigg{)}+2\bigg{(}{{\bf X}^{\prime}\cdot\ddot
{\bf X}\over 1-\dot {\bf X}^2}\bigg{)}^2+2\epsilon^2\bigg{(}{\dot{\bf
X}\cdot\ddot{\bf X} \over 1 -\dot{\bf X}^2}\bigg{)}^2 \cr
&+3\epsilon\dot\epsilon\bigg{(}{\dot{\bf X} \cdot\ddot{\bf X}\over 1-\dot{\bf
X}^2}\bigg{)} -{3\dot\epsilon\over\epsilon} \bigg{(}{\dot{\bf X}\cdot{\bf
X}^{\prime\prime}\over 1 -\dot{\bf X}^2}\bigg{)}\bigg{\}} \,.}\eqno(\new) 
$$  
The expressions for ${\bf f}^{\rm self}$ and $f^{0,\rm self}$, facilitate the
well-known renormalisation of the equations of motion (\tempeom) and string
energy, \eqnam{\reneom}  $$ \mu(\Delta)\bigg{(}\ddot{\bf X}-{1\over
\epsilon}\bigg{(}{{\bf X}^{\prime}\over\epsilon}\bigg{)}^{\prime}\bigg{)} =
{\bf f}^{\rm rad}\,,\quad \mu(\Delta)\dot\epsilon = f^{0,\rm
rad}\,,\quad E=\mu(\Delta)\int d\sigma\epsilon\,,\eqno(\new) $$  with the
renormalised string tension $\mu(\Delta)=\mu_0+2\pi\fa^2\log(\Delta/\delta)$,
while the expressions for ${\bf f}^{\rm rad}$  and $f^{0,\rm rad}$ represent
the finite radiation backreaction force in what we denote as the `local
backreaction approximation'. The assumptions underlying (\reneom)  are that
the dominant contribution to the integrals (\bsoln), (\lwpot) and
(\lwpotselfrad) come from string segments close to the point under
consideration and that the long-range fields are uncorrelated beyond the
string radius of curvature.  We believe these are reasonable assumptions for
general physical situations, notably for strings in a realistic interacting
network.

\sectbegin{3~}{Modified string equations of motion}

\noindent The equations of motion for the radiation force (\tempforce)  and
(\reneom) are rather complicated. However, at least for the string solutions
considered in ref.\refto{BS}, that is, closed loops and long periodic strings, 
we can demonstrate that some of these terms are sub-dominant (this has also
been verified numerically). In particular, it is possible to approximate
$H^{\rm rad}_{\mu\alpha\beta}$, ${\bf f}^{\rm rad}$ and $f^{0,\rm rad}$ by
\eqnam{\approxforce}  
$$ 
\eqalign{H^{\rm
rad}_{\mu\alpha\beta}&\approx-{2\fa\Delta\over 3\dot
X^4}X^{\tdot}_{[\rho}V_{\alpha\beta]}\,,\cr 
{\bf f}^{\rm
rad}&\approx{4\pi\fa^2\Delta\over 3}\bigg{[}\epsilon{\bf X}^{\tdot}- {1\over
\epsilon} \bigg{(}{{\bf X}^{\prime}\cdot{\bf X}^{\tdot}\over 1-\dot {\bf
X}^2}\bigg{)}{\bf X}^{\prime}\bigg{]}\,,\cr 
f^{0,\rm rad}&\approx{4\pi\fa^2\Delta\over
3}\bigg{[}{\epsilon^2\dot {\bf X}\cdot{\bf X}^{\tdot}\over 1-\dot {\bf X}^2}\bigg{]}
\,.}\eqno(\new) 
$$ 

This simplified form of the equations of motion using (\approxforce) still has
serious shortcomings because of the presence of the ${\bf X}^{\tdot}$ term. 
The equations  have unphysical, exponentially growing or `runaway' solutions
which will, for example, plague any potential numerical applications. 
Furthermore, one would be required to store information at three different
timesteps, fundamentally changing the nature of a numerical algorithm. It
appears, however, that both these problems can be circumvented by
resubstituting the equations of motion, that is, making the approximations
$\ddot {\bf X} = {\bf X}''/\epsilon^2$ and $ {\bf X}^{\tdot} = {\bf \dot
X}''/\epsilon^2$ in (\approxforce).  The equations of motion then acquire an
analogue of a viscosity term for which there are only damped solutions.

The reason for the suppression of the exponentially growing solution becomes apparent
if we consider simplified one-dimensional equations, 
\eqnam{\simpeqn} 
$$ \ddot X - X^{\prime\prime} = \alpha X^{\tdot}\,,\quad\longrightarrow\quad \ddot X
-  X^{\prime\prime} \approx \alpha \dot X^{\prime\prime} \,,\eqno(\new) 
$$  
where we have performed the resubstitution assuming that $\alpha$ is small.
We now take an approximately periodic solution,   
$X^{\prime\prime}\approx -\Omega^2X$, and we substitute the ansatz $X\sim e^{mt}$.
The solutions for (\simpeqn) are given respectively by the roots of the
following polynomials in $m$,  
\eqnam{\poly} $$f(m)=\alpha
m^3-m^2-\Omega^2\,,\quad\longrightarrow\quad
g(m)=-m^2-\alpha\Omega^2m-\Omega^2\,.\eqno(\new)$$
 If we rewrite $f(m)=(m^2+Am+\Omega^2+B)(\alpha m-C)$, then we see 
that $A=\alpha\Omega^2+{\cal O}(\alpha^2)$, $B={\cal O}(\alpha^2)$ and
$C=1+{\cal O}(\alpha^2)$. If we ignore terms ${\cal O}(\alpha^2)$, then the
solutions of $g(m)=0$ are approximately solutions of $f(m)=0$. However, the
real positive solution of $f(m)=0$, corresponding to the exponentially growing
solution, is not a solution of $g(m)=0$.
 
It remains to recast the simplified resubstituted equations of motion into a
first-order form accessible to numerical solution.  Defining ${\vec\alpha}={\bf
X}^{\prime}-\epsilon\dot {\bf X}$ and ${\vec\beta}={\bf X}^{\prime}+\epsilon\dot
{\bf X}$, the equations of motion can be rewritten as \eqnam{\numericaleom} 
$$
\eqalign{\mu(\Delta)\bigg{[}\dot{\vec\alpha}+\bigg{(}{{\vec\alpha}
\over\epsilon}\bigg{)}^{\prime}\bigg{]}&=-{1\over 2}\bigg{(}\epsilon{\bf
f}^{\rm rad}+f^{0,\rm rad}\dot {\bf X}\bigg{)}\,,\cr
\mu(\Delta)\bigg{[}\dot{\vec\beta}-\bigg{(}{{\vec\beta}\over\epsilon}
\bigg{)}^{\prime}\bigg{]}&={1\over 2}\bigg{(}\epsilon{\bf f}^{\rm
rad}+f^{0,\rm rad}\dot {\bf X}\bigg{)}\,,\cr \mu(\Delta)\dot\epsilon =
f^{0,\rm rad}\,,~~&\qquad \dot{\bf X} = {1\over
2\epsilon}\bigg{(}\vec\beta-\vec\alpha\bigg{)}\,.}\eqno(\new) 
$$  
Using the above, one
can then evolve string trajectories simply by modifying a total variation
non-increasing (TVNI) algorithm\refto{sod,ASa} which has already been
well-tested for string network evolution in an expanding universe. 

\sectbegin{4~}{Numerical and analytic comparisons}

\figure{decay.eps}{3.5in}{0.45in}{1}{Decay of $\varepsilon$ using the radiative
backreaction force (dotted line) and numerical field
theory simulations (solid line) for (a) a sinusoidal perturbation, (b) a
helicoidal perturbation with unequal left- and right-moving amplitudes, and (c)
a pure left moving helicoidal perturbation. Note the excellent quantitative
agreement for all three cases.}

\noindent The power due to the radiation
backreaction force can be found by differentiating the expression for the energy in
(\reneom)  \eqnam{\power} 
$$
P = - {dE\over dt}=
-{4\pi\fa^2\Delta\over 3}\int d\sigma {\epsilon^2 \dot {\bf X}\cdot{\bf
X}^{\tdot}\over 1 - \dot {\bf X}^2}\,.\eqno(\new)
$$ 
For closed loops of length
$L$, we take the integration over the range $0<\sigma<L$. Since the loop
oscillates relativistically with a period $T=L/2$, one can estimate $\dot {\bf
X}\sim {\cal O}(1)$, ${\bf X}^{\tdot}\sim {\cal O}(L^{-2})$. If $\Delta\sim
L$, then power per unit length is proportional to $L^{-1}$, which recovers the
well known result that the power loss from a loop is independent of its size
$L$. 

More caution has to be exercised in comparisons with long string configurations
because of the possibility of periodicity or other global correlations
interfering and suppressing radiation power. 
The generic result for a long string solution parameterized by  its
wavelength $L$ and amplitude to wavelength ratio $\varepsilon=2\pi A/L$,
where $A$ is the amplitude, is $\dot {\bf X}\sim {\cal O}(\varepsilon)$,
${\bf X}^{\tdot}\sim {\cal O}(\varepsilon L^{-2})$. Therefore the power per
unit length is $dP/dl =\beta\varepsilon^2/L$. The effect of this  power loss
can be calculated by analogy to the simple backreaction model presented in
ref.\refto{BS};  it will lead to an exponential decay of the amplitude and
oscillation energy per unit length,  \eqnam{\newmodel}
$$
\varepsilon=\varepsilon_0 \exp\bigg{(}-{\beta t\over 2\alpha\mu
L}\bigg{)}\,,\quad E/L = \mu+\alpha\mu\varepsilon_0^2 \exp\bigg{(}-{\beta
t\over\alpha\mu L}\bigg{)}\,.\eqno(\new)
$$ 
where $\alpha$, $\beta$ are
constants dependent on the string configuration, defined as in
ref.\refto{BS}.  

This result is in agreement with the exponential decay already shown
to occur for general configurations with unequal left- and right-moving
modes\refto{BS}.  However, the periodic solutions discussed in ref.\refto{Sak,BS}
were shown to have a weaker power law fall-off, $dP/dl =
\beta\varepsilon^4/L$, and some pure left-moving (or right-moving) solutions are known
to propagate indefinitely along a straight string solution with no
decay\refto{Vac}.   The resolution of this apparent contradiction lies in 
noticing that the renormalization procedure ignores any global
correlations of the long-range fields outside the cut-off scale $\Delta$.
For example, the non-radiating left-moving solution has an accompanying
left-moving perturbation in the long-range Goldstone field which is (unphysically)
correlated out to infinity\refto{Vac}.  Accordingly, if we suppress the
Goldstone field correlations at the curvature radius of the string
perturbations, then we should
recover the exponential decay anticipated in (\newmodel).

We have extensively tested our `local backreaction approximation', using the modified
Nambu equations of motion (\numericaleom), by comparing with field theory simulations
of  radiating strings in the Goldstone model (\goldaction) (see ref.\refto{BS}). 
Here, we shall only briefly report on the remarkable correspondence of the radiative
decay results, leaving  further details for a longer publication\refto{BSa}.  Note
that in the field theory simulations we evolve initial trajectories for which
Goldstone field correlations are suppressed by gaussian smoothing for
transverse distances greater than the string curvature radius. The massless field of
a perturbed string approaches that of a straight string at large distances from the
core, as we would expect in a general physical context for random string small-scale
structure. 
 
Fig.~1 illustrates the 
excellent quantitative agreement for three different long string configurations,
including a standing sinusoidal perturbation, a
helicoidal perturbation with unequal left- and right-moving amplitudes, and a
pure left-moving helicoidal perturbation.
In all three
cases, the decay is seen to be exponential.  The agreement persists for the
longest time for the (generic) unequal left- and right-moving configuration because
exponential decay is predicted even after field correlations have relaxed at
large distances.  By comparison to the simple backreaction
model for exponential decay (\newmodel) one can use the numerical field
theory results to estimate $\Delta\approx 0.1L$.  We had anticipated that
$\Delta$ should be normalized to a distance near the string radius of curvature $R$,
which for a sinusoidal perturbation is $R\sim L/4$.  The fact that $\Delta$ is
smaller than $R$ validates taking only the lowest order terms in
the expansion (\tempforce) for the radiative backreaction force.

Fig.~2 compares the evolution
of a sharp kink in the local backreaction approximation with a field theory
simulation. The results are almost indistinguishable except for the computational
advantages of the former which, in this case, saved a factor of $10^2$ in cpu time and
$10^4$ in allocated memory. Backreaction leads to a substantial rounding of the kink,
in agreement with the intuitive picture described in ref.\refto{BS}. A more thorough
investigation of this kink rounding, including spectral analysis, will be presented
in ref.\refto{BSa}.

\vdoublefigure{nam.eps}{field.eps}{1.5in}{0.8in}{1.2in}{1.5in}{2} {Decay of
kink perturbation $(\varepsilon_0$=$0.9)$ for using (a) the radiation
backreaction force and (b) numerical field theory. Notice the visible
rounding of the kink in both cases }

\sectbegin{5~}{Discussion and conclusions}

\noindent The local backreaction approximation which we have formulated
appears to provide a good description of radiative damping in a realistic
context, notably that of an evolving string network with uncorrelated left- and
right-movers.  In physical situations when the long-range string fields are
only correlated out to distances comparable with the string curvature radius,
numerical simulations confirmed the expected rate of exponential decay.  This
approach also appears to satisfactorily describe loop decay.  Of course, our
local approximation does not apply for solutions with global correlations in the
long-range fields, but our expectation is that such solutions are not of physical
relevance.   These  methods can also be applied to
gravitational and electromagnetic backreaction which we will discuss, along with 
more detailed tests, in a longer publication on this
subject\refto{BSa}.

These methods should prove useful for phenomenologically describing radiative effects
in simulations of string networks. The inclusion of such
backreaction terms may make tractable a number of outstanding cosmic string
problems, enabling us to understand the true nature and existence of the `scaling'
solution, as well as the amplitude and spectrum of axion and gravitational wave
backgrounds produced by radiating strings. Reformulated, this basic
approach to describing string radiation might also find application in
non-relativistic physical contexts, such as describing the radiation sound
waves by vortex-lines in condensed matter systems.

\nosectbegin{Acknowledgments}

\noindent We are grateful for helpful discussions with Brandon Carter,
Atish Dabholkar, Sun-Hong Rhie, David Bennett and Alex Vilenkin.  
The algorithm used for flat space Nambu string evolution was developed by EPS in
a collaborative project with Bruce Allen (see ref.\refto{ASa}).
We both acknowledge the support of the Science and Engineering Research
Council, in particular the Cambridge Relativity rolling grant (GR/H71550) and
Computational Science Initiative grants (GR/H67652 \& GR/H57585).



\def\hang{}


\def\jnl#1#2#3#4#5#6{\hang{#1 [#2], {\it #4\/} {\bf #5}, #6.}
									}



\def\prep#1#2#3#4{\hang{#1 [#2],  #4.}
									}

\def\proc#1#2#3#4#5#6{\hang{#1 [#2], in {\it #4\/}, #5, eds.\ (#6).}
}

\def\book#1#2#3#4{\hang{#1 [#2], {\it #3\/} (#4).}
									}



\def\prl{Phys.\ Rev.\ Lett.}
\def\pr{Phys.\ Rev.}
\def\pl{Phys.\ Lett.}
\def\np{Nucl.\ Phys.}

\def\cup{Cambridge University Press}

\def\skip{\vskip -4pt}

\nosectbegin{References}

\references

\baselineskip 10pt
\let\it=\nineit
\let\rm=\ninerm
\let\bf=\ninebf
\rm 

\refis{BS}
\prep{Battye, R.A., \& Shellard, E.P.S.}{1994}{Global string radiation}{to appear
in {\it \np} {\bf B} }
\skip

\refis{BSa}
\prep{Battye, R.A., \& Shellard, E.P.S.}{1994}{Radiative backreaction
on cosmic strings}{DAMTP Preprint}
\skip

\refis{ASa}
\jnl{Allen, B., \& Shellard, E.P.S.}{1990}{Cosmic string evolution---a
numerical simulation}{\prl}{64}{119} \proc{Shellard, E.P.S., \& Allen, B.}{1990}{On the evolution of 
cosmic strings}{Formation and Evolution of Cosmic Strings}{Gibbons, G.W.,
Hawking, S.W., \& Vachaspati, T.}{\cup} \skip
\skip

\refis{VV}
\jnl{Vilenkin, A., \& Vachaspati, T.}{1987}{Radiation of Goldstone bosons from
cosmic strings}{\pr}{D35}{1138}
\skip

\refis{KR}
\jnl{Kalb, M., \& Ramond, P.}{1974}{Classical direct interstring
action}{\pr}{D9}{2273}
\skip

\refis{Wit}
\jnl{Witten E.}{1985}{Cosmic Superstrings}{\pl}{153B}{243}
\skip

\refis{dirac}
\jnl{Dirac, P.A.M}{1938}{}{Proc. Roy. Soc. London}{A167}{148}
\skip

\refis{DSb}
\jnl{Davis, R.L., \& Shellard, E.P.S.}{1988}{Antisymmetric tensors and
spontaneous symmetry breaking}{\pl}{214B}{219}
\skip

\refis{DQ}
\jnl{Dabholkar, A., \& Quashnock, J.M.}{1990}{Pinning down the
axion}{\np}{B333}{815}
\skip

\refis{HCH}
\jnl{Copeland, E., Haws, D., \& Hindmarsh, M.B.}{1990}{Classical theory of radiating
strings}{\pr}{D42}{726}
\skip

\refis{barut}
\book{Barut, A.O.}{1964}{Electrodynamics and classical theory of fields
and particles}{Macmillan, New York}
\skip

\refis{jackson}
\book{Jackson, J.D.}{1975}{Classical Electrodynamics}{Wiley, New York}
\skip

\refis{sod}
\book{Sod, G.A.}{1985}{Numerical methods in fluid dynamics}{\cup}
\skip

\refis{Sak}
\jnl{Sakellariadou, M.}{1991}{Radiation of Nambu--Goldstone bosons from
infinitely long cosmic strings}{\pr}{D44}{3767}
\skip

\refis{Vac}
\jnl{Vachaspati, T.}{1986}{Gravitational effects of
cosmic strings}{\np}{B277}{593}  \skip

\endreferences

\vfill
\end